\documentclass[english,prl,twocolumn,superscriptaddress,showpacs,letterpaper]{revtex4}

\usepackage{ae} 
\usepackage[T1]{fontenc}
\usepackage[ansinew]{inputenc}
\usepackage{amsmath}
\usepackage{amssymb}
\usepackage{graphicx}
\usepackage{color}
\usepackage{epsfig}
\usepackage{booktabs}

    \def\be{\begin{eqnarray}}
    \def\ee{\end{eqnarray}}
    \def\no{\nonumber}

    \def\bn{\begin{enumerate}}
    \def\en{\end{enumerate}}
    \def\bi{\begin{itemize}}
    \def\ei{\end{itemize}}

    \def\({\left(\!}
    \def\){\!\right)}
    \def\<{\left\langle\,}
    \def\>{\, \right\rangle}
    \def\[{\left[}
    \def\]{\right]}

    \def\tilde{\widetilde}

    \def\s{\sigma}

    \def\p{\phi}

    \def\CN{{\cal N}}

    \def\p{\partial}

    \def\xpm{ x^{\pm} }
    \def\xp{ x^{+} }
    \def\xm{ x^{-} }
    \def\ypm{ y^{\pm}}
    \def\yp{ y^{+}}
    \def\ym{ y^{-}}

\usepackage{varioref}
\usepackage{makeidx}
\makeindex

\usepackage[english]{babel}
\begin{document}


\title{Minimal solution of the AdS/CFT crossing equation}

\author{Dmytro Volin}
\affiliation{Institut de Physique Th\'eorique, CNRS-URA 2306, C.E.A.-Saclay, F-91191 Gif-sur-Yvette, France \&\\ Bogolyubov
Institute for Theoretical Physics, 14b Metrolohichna Str.  \\
Kyiv, 03143 Ukraine}

\begin{abstract}
We solve explicitly the crossing equation under sufficiently general assumptions on the structure of the dressing phase. We obtain the BES/BHL dressing phase as a minimal solution of the crossing equation and identify the possible CDD factors.
\end{abstract}

\maketitle

\section{Introduction}
There is evidence that the spectral problem which appears in the context of the AdS/CFT correspondence can be solved exactly under the assumption of integrability \cite{MZ,BPR,BKS}. In analogy with the bootstrap approach in the relativistic theories \cite{Z}, an important object to identify is a factorized scattering matrix \cite{Staudacher:2004tk}. The knowledge of the scattering matrix leads to the solution of the spectral problem of the theory in a large volume in terms of the asymptotic Bethe Ansatz proposed in \cite{BS,B}.
From the symmetries of the theory it is possible to fix the scattering matrix up to a scalar factor \cite{B,AFZ} known also as dressing factor.

The necessity of the nontrivial dressing factor was first observed in \cite{Arutyunov:2004vx}. Later it was shown that it is constrained by the crossing equations \cite{Janik:2006dc}. A proposal for the logarithm of the scalar factor, known as the BES/BHL dressing phase, was given in \cite{Beisert:2006zy,BHL,BES}. This proposal passed many nontrivial checks. In particular it has a correct strong coupling asymptotics which was found from the algebraic curve solution of the string sigma model at tree level \cite{Arutyunov:2004vx} and at one loop \cite{Beisert:2005cw,Hernandez:2006tk,Gromov:2007cd}. The conjectured dressing phase was an important ingredient for the so called BES equation \cite{ES,BES}. The solution of the BES equation at week coupling \cite{BES} is in agreement with an explicit four-loop perturbative calculations in $\CN=4$ SYM \cite{BCDKS}. The strong coupling solution of the BES equaion \cite{Basso:2007wd,Kostov:2008ax} is in agreement with a two-loop string prediction \cite{Roiban:2007dq}. Now it is widely accepted that the BES/BHL dressing phase gives a correct scalar factor of the AdS/CFT scattering matrix.

In integrable relativistic field theories the solution of the crossing equation is not unique. The correct scalar factor is usually a "minimal" solution of the crossing equation, \textit{i.e} the solution with a minimal number of the singularities in the physical strip \cite{Z}. All other solutions of the crossing equation can be obtained via multiplication by the CDD factors.

In this note we argue that this is the case also for the AdS/CFT integrable system. We explicitly solve the crossing equation and show that the "minimal" solution coincides with the BES/BHL dressing factor. We also comment on the possible form of the CDD factors.

\section{Crossing relation at the mirror point}
In the following we will use the Jukowsky map $x[u]$ defined by
\be
\label{Jukowsky}
  x=\frac{u}{2g}\(1+\sqrt{1-\frac {4g^2}{u^2}}\),\ \ \frac {u}{g}=x+\frac 1x.
\ee
  We introduce the following shorthand notations. By $x,y,$ and $z$ we denote respectively the images of $u,v,$ and $w$. We also denote $\xpm\equiv x[u\pm i/2]$ and $\ypm\equiv x[v\pm i/2]$. We take the branch of the Jukowsky map such that $|x|>1$ if opposite is not mentioned explicitly.

The crossing equation is formulated as follows \cite{Janik:2006dc,Arutyunov:2006iu}:
\be\label{crossing}
  \sigma[u,v]\sigma^{cross}[u ,v]=\frac{\ym}{\yp}\frac{\xm-\yp}{\xp-\yp}\frac{1-\frac 1{\xm\ym}}{1-\frac 1{\xp\ym}},
\ee
where $\sigma[u,v]$ is a dressing factor and $\sigma^{cross}[u,v]$ is its crossing transform.

The dressing factor is a multivalued function of $u$ and $v$. In the following we will fix $v$ and consider $\s$ as a function of $u$.

\begin{figure}[t]\label{fig:crossing}
\centering
\includegraphics[width=6.00cm]{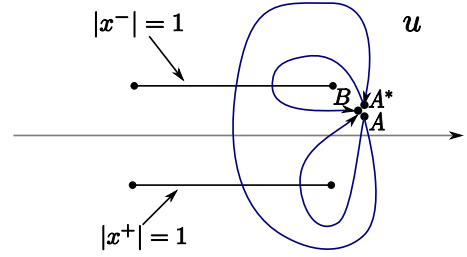}
\caption{The crossing $A^*$ and the mirror points $B$ in the $u$-plane.}
\end{figure}

The crossing transformation acts as the analytical continuation of $\sigma$ along the contour $AA^*$ in the $u$-plane, depicted in Fig.~1, which encircles the branch points $2g\pm i/2$. As it is seen from (\ref{crossing}), the dressing factor $\s$ has a nontrivial monodromy along the contour $AA^*$.

It is possible to write the crossing equation as a periodicity condition using the elliptic uniformization \cite{Janik:2006dc}. In terms of the elliptic variable $s$ used in \cite{KSV1} and related to one in \cite{Janik:2006dc} by a Gauss-Landen transformation the crossing transformation is given by
$\s^{cross}[s]=\s[s+2iK']$. Since $\s[s+8iK']=\s[s]$, the branch points $u=\pm 2g\pm i/2$ may be of the fourth or the second order. We assume in the following that they are of square root type which is compatible with the analytical structure of the Bethe Ansatz equations \cite{BS}.

The dressing factor $\sigma[u,v]$ can be  represented in the following form \cite{Arutyunov:2004vx,Arutyunov:2006iu}:
\be\label{param}
\s[u,v]&=&e^{i\theta[u,v]},\\
\theta[u,v]&=&\chi[\xp,\ym]-\chi[\xm,\ym]-\chi[\xp,\yp]+\chi[\xm,\yp],\no
\ee
where $\chi[x,y]$ has the symmetry $\chi[x,y]=-\chi[y,x]$. To fix the solution of (\ref{crossing}) we assume that $\chi$ is analytic single-valued function for $|x|>1$. Since the Jukowsky map (\ref{Jukowsky}) resolves the simple branch points of $\sigma[u,v]$ at $u=\pm 2g\pm i/2$, the function $e^{i\chi[x,y]}$ is meromorphic in the vicinity of the points $x=\pm 1$.

The crossing transformation acts on the functions $\xpm$ by inverting them: $\xpm\to 1/\xpm$. Since $\chi[x,y]$ may not be single-valued in the domain $|x|<1$, the function $\sigma^{cross}$ will be written in terms of the functions $\chi[1/x,y]$ defined on different Riemann sheets of the $x$ plane. To avoid this problem we analytically continue the crossing equation (\ref{crossing}) along the path $A^*B$ to the region with mirror kinematics.
The crossing equation at the point $B$ is written as
\be\label{mirrorcros}
  \s^{A^*B}[u,v]\s^{AB}[u,v]=\frac{1-\frac 1{\xp\yp}}{1-\frac 1{\xm\ym}}\frac{1-\frac 1{\xm\yp}}{1-\frac 1{\xp\ym}},
\ee
where $A^*B$ and $AB$ denote the paths which were used for analytical continuation of $\s[u,v]$.

 The point $B$ is not necessarily at the same position with the points $A$ and $A^*$ in the $u$ plane. The position of the point $B$ can be chosen in a way that $AB$ and $A^*B$ correspond to the shifts of the $s$ variable $s\to s\pm iK'$. These shifts relate the theory with its mirror \cite{Arutyunov:2007tc}.

In the following we will use the functions $\s_1[x,v]$ and $\s_2[x,y]$ given by
\be
  \s_1[x,v]=e^{i\chi[x,\ym]-i\chi[x,\yp]},\ \
  \s_2[x,y]=e^{i\chi[x,y]}.
\ee

Since the paths $A^*B$ and $AB$ cross only one simple branch cut, we are allowed to write \footnote{Strictly speaking, we also have to demand absence of the branch points between the cuts $|\xm|=1$ and $|\xp|=1$. This demand is analogic to the demand of analyticity of the $S$-matrix in relativistic field theories.}
\be\label{ABA}
  \sigma^{A^*B}=\frac {\s_1[1/\xp,v]}{\s_1[\xm,v]},\ \
  \sigma^{AB}=\frac {\s_1[\xp,v]}{\s_1[1/\xm,v]}.
\ee
Now all four functions $\s_1$ which are used in (\ref{ABA}) are on the same Riemann sheet of the $x$ plane.

For the further analysis it will be convenient to write the crossing equation (\ref{mirrorcros}) in terms of the shift operator
\be
  D\equiv e^{\pm \frac i2\p_u}:f[u]\mapsto f[u\pm i/2],
\ee
so that $\xp=D\ x$ and $\xm=D^{-1}\ x$. Namely, introducing the notation
\be
  f^{D^{\pm 1}}\equiv e^{D^{\pm 1}\log[f]},
\ee
we can write (\ref{mirrorcros}) as
\be\label{mirror}
  &&\(\ \s_1[x,v]\s_1[1/x,v]\ \)^{D-D^{-1}}=\(\frac{x-\frac 1{\yp}}{x-\frac 1{\ym}}\)^{\!\!D+D^{-1}}.
\ee
 The shift operator $D$ is not well defined inside the strip $|{\rm Re}[u]|\leq 1$ since we can cross the cut of $x[u]$ and go to another sheet. To avoid this ambiguity we will consider the crossing equation (\ref{mirror}) outside this strip, solve it, and then analytically continue the solution.
\section{Solution}
The function $\s_1[x,v]\s_1[1/x,v]$ as a function of $u$ does not have a branch cut $[-2g,2g]$. A solution of (\ref{mirror}) with this property is given by
\be\label{s1}
  &&\s_1[x,v]\s_1[1/x,v]=\(\frac{x-\frac 1{\yp}}{x-\frac 1\ym}\)^{-\frac{D^2}{1-D^2}+\frac {D^{-2}}{1-D^{-2}}},\\
  &&\frac{D^{\pm 2}}{1-D^{\pm 2}}=D^{\pm 2}+D^{\pm 4}+\ldots\ .\no
\ee
Strictly speaking, this expression should be regularized to have a precise meaning. However, the regulating terms will cancel for the complete dressing factor $\s[u,v]$.

The expression (\ref{s1}) can be further simplified using the fact that $\s_1[x,v]=\s_2[x,\yp]/\s_2[x,\ym]$:
\be
  \s_2[x,y]\s_2[1/x,y]=\(\frac{x-\frac 1y}{\sqrt{x}}\)^{-\frac{D^2}{1-D^2}+\frac {D^{-2}}{1-D^{-2}}}.
\ee
The multiplier $1/\sqrt{x}$ does not contribute to (\ref{s1}). It is needed for the consistency with the antisymmetry of $\chi[x,y]$ with respect to interchange $x\leftrightarrow y$. Indeed, a direct calculation shows that
\be\label{s2}
  \s_2[x,y]\s_2[1/x,y]\s_2[x,1/y]\s_2[1/x,1/y]=\no\\=(u-v)^{-\frac{D^2}{1-D^2}+\frac {D^{-2}}{1-D^{-2}}}=\frac{\Gamma[1+i(u-v)]}{\Gamma[1-i(u-v)]},
\ee
whose logarithm is antisymmetric with respect to $u\leftrightarrow v$ as it should.

The ratio of gamma-functions resembles the solutions of the crossing equation in the relativistic theories. Note that the typical dressing factor in the relativistic theories is given by an expression of the type \cite{Volin06}
\be
  \theta^{-\frac {D^2}{1+D^2}+\frac {D^{-2}}{1+D^{-2}}},
\ee
\textit{i.e} with the opposite sign in the denominator.

By taking the logarithm of (\ref{s2}) we get a simple Riemann-Hilbert problem which is solved by
\be\label{s2solution}
  \chi[x,y]=-i\tilde{K}_u\tilde{K}_v\log\[\frac{\Gamma[1+i(u-v)]}{\Gamma[1-i(u-v)]}\],
\ee
with the kernel $\tilde K$ defined by
\be
  (\tilde K\cdot F)[u]=\int_{-1+i0}^{1+i0}\frac{dw}{2\pi i}\frac {x-\frac 1x}{z-\frac 1{z}}\frac 1{w-u}F[w].
\ee
The kernel $\tilde K$ is constructed to satisfy the following equation:
\be\label{Cauchy}
  (\tilde K\cdot F)[u+i0]+(\tilde K\cdot F)[u-i0]=F[u],\  u^2<4g^2.
\ee
The subscripts $u$ and $v$ in (\ref{s2solution}) refer to action of $\tilde K$ on $u$ and $v$ variables respectively.

This solution was chosen among the other possible solutions by the requirement that $\chi[x,y]$ should be analytic for $|x|>1$ and $\chi[x,y]\to{\rm const},\ x\to\infty$.

The expression (\ref{s2solution}) can be rewritten in the form proposed by Dorey, Hofman, and Maldacena \cite{DHM} if to rewrite the action of the kernel $\tilde K$ as an integral in the Jukowsky plane
\be\label{toDHM}
  (\tilde K\cdot F)[u]=\oint_{|z|=1}\frac {dz}{2\pi i}\frac 1{x-z}F\[g\(z+\frac 1z\)\]\!\!+\!\rm const
\ee
and to note that the constant term does not contribute to the dressing phase.
This kind of transformations was discussed in \cite{Kostov:2008ax}.

Therefore we see that solution we obtained is nothing but the BES/BHL dressing phase.

The solution (\ref{s2solution}) of the crossing equation is not unique. It can be thought as a minimal solution in the sense that we chose the solution with a minimal possible number of singularities.  The form of the CDD factors can be strongly constrained assuming the decomposition (\ref{param}), square root type of the branch points and absence of the branch points different from the ones that are present in the BES/BHL dressing phase. Then the CDD factor should satisfy the equation
\be
  (\s_{1,{\rm CDD}}[x,v]\s_{1,{\rm CDD}}[1/x,v])^{D-D^{-1}}=1.
\ee
We see that the function $f_{\rm CDD}[u]=\s_{1,\rm CDD}[x,v]\s_{1,\rm CDD}[1/x,v]$ should be periodic with the period $i$. Since by construction $f_{\rm CDD}[u]$ does not have a branch cut $[-2g,2g]$, due to periodicity it cannot have other branch cuts as well. Therefore $f_{\rm CDD}$ is a meromorphic function of $u$ and $\s_{1,\rm CDD}[x,v]$ is a meromorphic function of $x$. As a consequence, all the branch points of the dressing factor are resolved by introducing an elliptic parametrization. An arbitrary CDD factor is therefore a meromorphic function on the torus which satisfies the condition
\be
  \s_{\rm CDD}[s]\s_{\rm CDD}[s+2iK']=1.
\ee

\section{Analytical structure}
The investigation of the analytical structure of the solution (\ref{s2solution}) is based on the property (\ref{Cauchy}). It is instructive to write (\ref{s2solution}) as
\be\label{repre}
 -i \chi=\tilde K_u\(\frac {D^2}{1-D^2}-\frac {D^{-2}}{1-D^{-2}}\)\tilde K_v \log[u-v].
\ee

Let us consider $\chi$ as function of $u$. We will keep the notation $\chi$ to denote $\chi[x[u],y]$ in the domain $|x|>1$. In this domain $\chi$ is analytic everywhere except on the Jukowsky cut $|x|=1$. Crossing this cut will bring us to the other Riemann sheet. We will denote the function $\chi$ on this sheet by $\chi'$. From (\ref{Cauchy}) and (\ref{repre}) we deduce that
\be\label{secondsheet}
  -i \chi'=i\chi+\sum_{n\neq 0}{\rm sign}[n]\tilde K_v\log[u-v+in].
\ee
We see that $\chi'$ has infinite set of simple branch cuts given by the condition $|x[u+in]|=1$. By construction $\chi'$ is defined in the domain $|x[u]|<1$ and $|x[u+in]|>1$ for $n\neq 0$.

Passing through one of the cuts $|x[u+im]|=1$ with $m\neq 0$ brings us to yet a new Riemann sheet which is defined by $|x|<1$ and $|x[u+im]|<1$. The corresponding function $\chi^{''}_{m}$ is given by
\be\label{thirdsheet}
  -i\chi^{''}_{m}=i\chi'+{\rm sign}[m]\log[u-v+i m].
\ee
The second term in the  in the r.h.s of (\ref{thirdsheet}) leeds to the DHM poles described in \cite{DHM}. These poles should be squared in the scattering matrix since the dressing factor contributes as $\s^2$.
\begin{figure}[t]\label{fig:crossing}
\centering
\includegraphics[width=6.00cm]{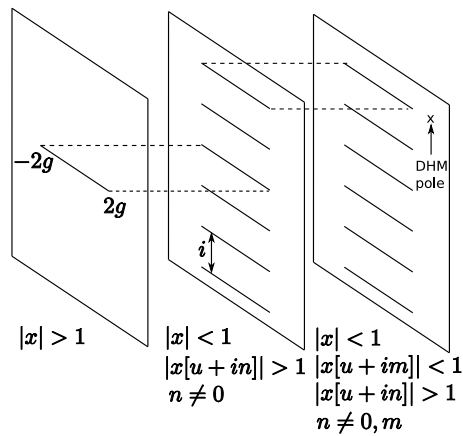}
\caption{Analytical structure of $\sigma_2[x,y]$ as a function of $u$.}
\end{figure}

The analytical properties of the dressing phase follow from the analytical properties of the function $\chi$ and the decomposition (\ref{param}). Note however that from the crossing equation (\ref{crossing}) it follows that on the Riemann sheet which contains the point $A^*$ the dressing phase should have only two cuts $|\xpm|=1$. The cancelations of the other cuts \footnote{These cancelations were also independently observed by N.Gromov and P.Vieira} on this sheet between $\s_2[\xp,y]$ and $\s_2[\xm,y]$ can be easily seen from (\ref{secondsheet}).

\

The representation (\ref{repre}) is closely related to the representation for the dressing kernel in \cite{KSV1,Volin:2008kd}. As it shown in \cite{Volin:2008kd}, the dressing kernel can be nontrivially simplified at strong coupling. The integral representation of the dressing kernel can be also obtained by an inverse half-Fourier transform \cite{Kostov:2008ax} of the dressing kernel of the BES equation \cite{BES}.

\ \\

{\bf Note added}

\noindent
After the work on this project was already finished the paper \cite{Arutyunov:2009kf} appeared. In \cite{Arutyunov:2009kf} it was shown that the BES/BHL dressing phase satisfies the crossing relation (\ref{crossing}). In opposite to \cite{Arutyunov:2009kf} here we derive the dressing phase in a constructive way. Also in \cite{Arutyunov:2009kf} the analytic structure of the dressing phase as a function on the torus is discussed. Here we present the analytic structure of the dressing phase as a function of $u$.

\ \\

{\bf Acknowledgments}

\noindent
The author thanks to I.Kostov and D.Serban for useful discussions and to S.Frolov for useful comments on the article.
This work has been supported by the  European Union
through ENRAGE network (contract MRTN-CT-2004-005616).


\begin{thebibliography}{99}

\bibitem{MZ}
J.~A.~Minahan and K.~Zarembo,
\textsf{JHEP~0303,~013~(2003)}, \texttt{{hep-th/0212208}}.
\bibitem{BPR}
  I.~Bena, J.~Polchinski and R.~Roiban,
  Phys.\ Rev.\  D {\bf 69}, 046002 (2004)
  [arXiv:hep-th/0305116].
\bibitem{BKS}
N.~Beisert, C.~Kristjansen and M.~Staudacher,
 \textsf{{Nucl.~Phys.~B664,~131~(2003)}},
\texttt{{hep-th/0303060}}.
\bibitem{Z}
  A.~B.~Zamolodchikov and A.~B.~Zamolodchikov,
  Annals Phys.\  {\bf 120}, 253 (1979).
\bibitem{Staudacher:2004tk}
  M.~Staudacher,
  JHEP {\bf 0505}, 054 (2005)
  [arXiv:hep-th/0412188].
\bibitem{BS}
  N.~Beisert and M.~Staudacher,
  Nucl.\ Phys.\  B {\bf 727}, 1 (2005)
  [arXiv:hep-th/0504190].
\bibitem{B}
  N.~Beisert,
  Adv.\ Theor.\ Math.\ Phys.\  {\bf 12}, 945 (2008)
  [arXiv:hep-th/0511082].

\bibitem{AFZ}
 G.~Arutyunov, S.~Frolov and M.~Zamaklar,
  JHEP {\bf 0704}, 002 (2007)
  [arXiv:hep-th/0612229].


\bibitem{Arutyunov:2004vx}
  G.~Arutyunov, S.~Frolov and M.~Staudacher,
  JHEP {\bf 0410}, 016 (2004)
  [arXiv:hep-th/0406256].

\bibitem{Janik:2006dc}
  R.~A.~Janik,
  Phys.\ Rev.\  D {\bf 73}, 086006 (2006)
  [arXiv:hep-th/0603038].

\bibitem{Beisert:2006zy}
  N.~Beisert,
  Mod.\ Phys.\ Lett.\  A {\bf 22}, 415 (2007)
  [arXiv:hep-th/0606214].

\bibitem{BHL}
  N.~Beisert, R.~Hernandez and E.~Lopez,
  JHEP {\bf 0611}, 070 (2006)
  [arXiv:hep-th/0609044].

\bibitem{BES}
  N.~Beisert, B.~Eden and M.~Staudacher,
   \textsf{J.\ Stat.\ Mech.\ {\bf 0701}, P021 (2007)},
  \texttt{hep-th/0610251.}

\bibitem{Beisert:2005cw}
  N.~Beisert and A.~A.~Tseytlin,
  Phys.\ Lett.\  B {\bf 629}, 102 (2005)
  [arXiv:hep-th/0509084].

\bibitem{Hernandez:2006tk}
  R.~Hernandez and E.~Lopez,
  JHEP {\bf 0607}, 004 (2006)
  [arXiv:hep-th/0603204].

\bibitem{Gromov:2007cd}
  N.~Gromov and P.~Vieira,
  Nucl.\ Phys.\  B {\bf 790}, 72 (2008)
  [arXiv:hep-th/0703266].

\bibitem{ES}
  B.~Eden and M.~Staudacher,
  J.\ Stat.\ Mech.\  {\bf 0611}, P014 (2006)
  [arXiv:hep-th/0603157].

\bibitem{BCDKS}
Z.~Bern, M.~Czakon, L.~J.~Dixon, D.~A.~Kosower and V.~A.~Smirnov,
 \textsf{  Phys. Rev. D75 (2007) 085010}
\texttt{{hep-th/0610248}}.

\bibitem{Basso:2007wd}
  B.~Basso, G.~P.~Korchemsky and J.~Kotanski,
  Phys.\ Rev.\ Lett.\  {\bf 100}, 091601 (2008)
  [arXiv:0708.3933 [hep-th]].
\bibitem{Kostov:2008ax}
  I.~Kostov, D.~Serban and D.~Volin,
  JHEP {\bf 0808}, 101 (2008)
  [arXiv:0801.2542 [hep-th]].
\bibitem{Roiban:2007dq}
  R.~Roiban and A.~A.~Tseytlin,
  JHEP {\bf 0711}, 016 (2007)
  [arXiv:0709.0681 [hep-th]].



\bibitem{KSV1}
  I.~Kostov, D.~Serban and D.~Volin,
   \textsf{Nucl.\ Phys.\  B {\bf B789} (2008) 413},
   \texttt{arXiv:hep-th/0703031}.

\bibitem{Arutyunov:2006iu}
  G.~Arutyunov and S.~Frolov,
  Phys.\ Lett.\  B {\bf 639}, 378 (2006)
  [arXiv:hep-th/0604043].

\bibitem{Arutyunov:2007tc}
  G.~Arutyunov and S.~Frolov,
  JHEP {\bf 0712}, 024 (2007)
  [arXiv:0710.1568 [hep-th]].


\bibitem{Volin06}
  D.~Volin,
  arXiv:0904.2744 [hep-th].

\bibitem{DHM} N.~Dorey, D.~M.~Hofman and J.~M.~Maldacena,
  Phys.\ Rev.\  D {\bf 76}, 025011 (2007)
  [arXiv:hep-th/0703104].
  
\bibitem{Volin:2008kd}
  D.~Volin,
  arXiv:0812.4407 [hep-th].


\bibitem{Arutyunov:2009kf}
  G.~Arutyunov and S.~Frolov,
  arXiv:0904.4575 [hep-th].









\end{thebibliography}
\end{document}